\input epsf
\documentclass[a4paper,11pt]{article}
\usepackage{jcappub}
\usepackage{epsfig}

\def\bit{\begin{itemize}}
\def\eit{\end{itemize}}
\def\ben{\begin{enumerate}}
\def\een{\end{enumerate}}
\def\beq{\begin{equation}}
\def\eeq{\end{equation}}
\def\bea{\begin{eqnarray}}
\def\eea{\end{eqnarray}}
\def\bq{\begin{quote}}
\def\eq{\end{quote}}
\def \lsim{\mathrel{\vcenter
     {\hbox{$<$}\nointerlineskip\hbox{$\sim$}}}}
\def \gsim{\mathrel{\vcenter
     {\hbox{$>$}\nointerlineskip\hbox{$\sim$}}}}
\def\gappeq{\mathrel{\rlap {\raise.5ex\hbox{$>$}}
{\lower.5ex\hbox{$\sim$}}}}
\def\lappeq{\mathrel{\rlap{\raise.5ex\hbox{$<$}}
{\lower.5ex\hbox{$\sim$}}}}

\def\a{\alpha}
\def\b{\beta}

\begin{document}
\renewcommand{\thefootnote}{\fnsymbol{footnote}}
\begin{center}
{\Large {\bf 
Bose Einstein condensation of the classical axion field in cosmology?}}
\vskip 25pt
{\bf   Sacha Davidson \footnote{E-mail address: s.davidson@ipnl.in2p3.fr} and  Martin Elmer}\footnote{E-mail address: m.elmer@ipnl.in2p3.fr } 
 
\vskip 10pt  
{\it IPNL, Universit\'e de Lyon, Universit\'e Lyon 1, CNRS/IN2P3, 4 rue E. Fermi 69622 Villeurbanne cedex, France}\\
\vskip 20pt
{\bf Abstract}
\end{center}

\begin{quotation}
  {\noindent\small 
The axion is a motivated cold dark matter   candidate, which it
would be interesting to distinguish from weakly interacting
massive particles.
Sikivie has suggested that axions could behave differently during
non-linear galaxy evolution, if they form a Bose-Einstein condensate,
{ and argues that ``gravitational thermalisation'' drives them
to a Bose-Einstein condensate during the radiation dominated era. 
Using classical equations of motion during linear
structure formation, we explore whether  the gravitational
interactions of  axions can generate { enough} entropy.  }
At linear order in $G_N$, we interpret that the principle
activities of gravity are to expand the Universe and
grow density fluctuations. { To quantify the rate of entropy creation we use the anisotropic stress to
estimate a short dissipation scale for axions, } 
which does not confirm previous estimates of their
gravitational thermalisation rate.

\vskip 10pt
\noindent
}

\end{quotation}

\vskip 20pt  

\setcounter{footnote}{0}
\renewcommand{\thefootnote}{\arabic{footnote}}

\section{Introduction}
\label{intro}

Axions  \cite{rev,PQ,revggr} 
are  hypothetical light  pseudoscalar  bosons,
with phenomenological 
 and  theoretical
  attractions.  
They  could  constitute  a quarter of
the mass density of the Universe today, being
the cold dark matter(CDM) responsible for the
growth of galaxies and large scale structure.
  The axion  constitutes a  minimal
and theoretically  attractive  candidate,
because  it  arises in models
which  solve the ``strong CP problem'' of
QCD, and is accompanied by no
other new particles  at accessible
energies  \cite{DFS,russes}.

An interesting question is whether axions
can be distinguished from more massive 
CDM  candidates, such as weakly interacting
massive particles(WIMPs) \cite{WIMP}.
The axion $a$ is the Goldstone boson of the
Peccei-Quinn $U_{\rm PQ}(1)$ symmetry \cite{PQ} that  breaks
at a high scale $f_{\rm PQ} \gsim 4 \times 10^8$ GeV.   
It acquires a small mass $m_a \lsim 0.01$ eV  
by mixing with the pion. The very light
axion can nonetheless constitute CDM
if it is non-relativistic,
which requires a non-thermal production mechanism.
For instance,  an oscillating classical axion 
field can be produced at the QCD phase transition. 
It is well-known that the energy
density of  a homogeneous
and isotropic scalar field redshifts \cite{DineFischler,aCDM}
like CDM, and
that   the linear growth of
density fluctuations is the  same for
axions and  WIMPs \cite{linearaxions,HwangNoh,ratra,NambuSasaki}\footnote{The
doubts in the appendix of  \cite{K+S} are addressed
in \cite{KSS} and \cite{ratra,Ratra2}.}. Sikivie
and collaborators \cite{SY1,ESTY,ESTY2}  have extensively
explored the differences between
axions and WIMPs, in search of distinguishing
observables. In an extended study \cite{ESTY},
Erken, Sikivie, Tam and Yang
 (hereafter ESTY; see
also the formalised analysis of \cite{SY})
argue that axion-CDM forms a Bose-Einstein (BE) condensate due
to gravitational scattering at  photon
temperatures $\sim$ keV, and since 
the BE condensate can support vortices,  this  allows
caustics in the dark matter distribution
of our galaxy today.  
They  conclude  that axions behave differently from
WIMPs during non-linear structure formation.

The results of \cite{ESTY,SY}  are obtained using Quantum Field Theory
and Newtonian gravity, and   have various curious features,
which are mentioned at the end of section \ref{sec:rev}.
The aim of this paper is to study axion evolution in  an alternative 
formalism. We use classical equations of motion for
axions in an expanding Universe with metric perturbations\footnote{The
beautiful quantum analysis in a perturbed expanding Universe
of Nambu and Sasaki \cite{NambuSasaki} is helpful for
making contact between the quantum and classical studies.}.
A classical analysis  should give the lowest order solution, and gravity is
a classical theory. We study axion evolution in
the early Universe,  after
the QCD phase transition  and in the regime where departures from
the homogeneous and isotropic  solutions 
can  be treated in linear
perturbation theory. (The Appendix
recalls that   
the effect of the homogeneous and
isotropic  gravitational interactions  
 is to redshift the axion momenta.)  Even during this period, the equations
of motion for the axion field are non-linear,
so we study  instead the fluctuations of
the axion stress-energy tensor.  This is
the current to which gravity couples, so we
anticipate that its components are  appropriate
variables for describing
gravitational interactions  of axions.

Section \ref{sec:rev} contains a review of  axion
cosmology,  and some basics of Bose-Einstein condensation.
We wish to know if the  gravitational interactions
 { of   axions  generate entropy 
during linear structure
formation, so we are looking for 
a dissipative process.}
In section \ref{sec:Tmn}, we focus on physics inside the horizon,
and equate the axion stress-energy tensor
 of the perturbed Universe, with the stress-energy
tensor of an imperfect fluid. This gives 
an  estimate of  the  viscosity of the  axion fluid due
to metric/density fluctuations, and viscosity, in  fluid
dynamics, wipes out short distance modes. 
The damping scale this gives is very short, and does
not indicate that the axion field forms a BE condensate.
Section \ref{sec:disc} discusses our result
and  its relation to the literature, and 
section  \ref{sec:sum} is a summary.


\section{ Review }
\label{sec:rev}

After a brief introduction  of the theory of axions, 
 section \ref{ssec:a} tells the story of their cosmological evolution
before and after the epochs of interest in this paper. 
In section \ref{sec:BEC} we briefly
recall what is  Bose-Einstein condensation
in equilibrium and non-equilibrium field theory.

\subsection{Axions and their  cosmology }
\label{ssec:a}

The strong CP problem of QCD  is that
instantons should generate an 
 $F\widetilde{F}$  term in the Lagrangian,
with coefficient $\theta \sim 1$. However,
the non-observation of   the neutron electric
dipole moment \cite{nedm} implies $\theta \lsim 10^{-10}$
\cite{theta}. 
This discrepancy  can be  explained by making
$\theta$ a  massive dynamical field --- the axion \cite{PQ}.
Axion models can be constructed 
 by extending the particle content
of the Standard Model Lagrangian
to allow a global chiral U(1) symmetry,
referred to as a Peccei-Quinn \cite{PQ} symmetry,
which is broken by colour anomalies \cite{tH}.  The
$U_{\rm PQ}(1)$ is also  spontaneously broken at a high scale,
which  ensures that
the Goldstone boson is the only
new particle  at low energies, and
that it has tiny interactions
with the SM \cite{DFS,russes}. 
The   couplings to photons and  gluons,  are suppressed by
 $ 1/f_{\rm PQ}$; other
interactions
of axions can be found  in   \cite{revSred}.
The colour anomalies give this axion
a small mass from mixing with the pion:
\beq
m_a \simeq \frac{m_\pi f_\pi}{f_{\rm PQ}}  \frac{\sqrt{m_um_d}}{m_u+m_d}
 \simeq 6 \times 10^{-6} eV 
\frac{10^{12}  GeV}{f_{\rm PQ}} ~~~.
\label{mf}
\eeq

  Axions are searched for 
 in various experiments \cite{CAST,ADMX}, and can be
constrained by astrophysical observations \cite{revggr}.
The most stringent lower bound is $f_{\rm PQ} \gsim 4 \times 10^{8}$ GeV,
to ensure that axions do not carry to much energy
out of stars  \cite{revggr,RGnew}.

Due to the high scale  $f_{\rm PQ}$, it is unclear whether
the  Peccei Quinn (PQ) phase transition occurs before or after inflation. 
Both scenarios  have been extensively studied 
(for reviews and references, see, {\it e.g.} 
 \cite{mukhanov,BGBL,efstathiou}),
with emphasis on large scale  density fluctuations
relevant to the CMB and
galaxy formation.  The axion fluctuations we
will study in this paper are on
 distances  at least a million times shorter.

If  the PQ phase transition occurs sufficiently before
inflation,   a coherent  patch of  axion field will be
 inflated beyond the size of the visible Universe. 
During inflation, 
the  axion field, like the inflaton, 
will develop   fluctuations on  scales relevant
to the CMB and large scale structure:  
$\delta a/a \sim H_{inf}/(2 \pi f_{\rm PQ})$ \cite{mukhanov}.
Provided that  $ H_{inf} \ll f_{\rm PQ}$, these 
will be small fluctuations on a homogeneous and
isotropic axion field (see \cite{BGBL} for 
 the case $ H_{inf} \to f_{\rm PQ}$). 
When the axion acquires a potential at the QCD
phase transition,  it inherits the
adiabatic density perturbations 
that the inflaton imprinted on the plasma,
and in addition, its own  fluctuations become
isocurvature density perturbations \cite{BE}\footnote{For a
pedagogical introduction to adiabatic and isocurvature fluctuations, see 
{\it e.g.} \cite{Langlois}}. The
non-observation in CMB data of isocurvature
perturbations   allows to 
constrain  this axion scenario where the PQ
transition is before inflation \cite{BGBL}.

When the  PQ   phase transition occurs after inflation,
which is the scenario of interest in this paper, 
 the axion field will have random different values in different
causally connected volumes of the Universe. 
The coherence scale of the field  grows with  the
horizon  \footnote{This follows from
the equations of motion for a massless
field in the Friedman Roberson Walker Universe.}
until the QCD phase transition,
and  a network of
cosmic strings develops.   
The strings  disappear after the QCD
phase transition \cite{strings,DavisShellard},
radiating axions with  momentum of order
the Hubble expansion rate.
This population of non-relativistic 
axions is difficult to calculate
reliably \cite{strings,DavisShellard}; 
a recent estimate of their contribution    
to the CDM density today \cite{strings} is:
\beq
\Omega_a  \sim 0.2
\times 
\left(\frac{f_{\rm PQ}}{ 10^{11} ~{\rm GeV}}\right)^{6/5}
~~~~~{\rm (from ~strings).}
\label{Oastrings}
\eeq
In this paper, for simplicity, we neglect this
bath of cold axion particles. 
We also restrict to axion models which do not generate
 domain walls.

Until shortly before the QCD Phase Transition,
the potential for the axion  field was flat.
Afterwards, massive pions appear, and  the
axion field develops a potential(we follow here 
 \cite{rev})
\beq
V(a) \approx f^2_{\rm PQ}m_a^2 [1-\cos(a/f_{\rm PQ})] \simeq \frac{1}{2} m_a^2 a^2
-  \frac{1}{4!} \frac{m_a^2 }{f_{\rm PQ}^2} a^4 ~~.
\label{eqn3}
\eeq
 The QCD phase transition is  a cross-over
in lattice simulations \cite{QCDlattice}, 
suggesting that the turn-on of the axion mass
is  a smooth and homogeneous process.
A few Hubble times later, 
the mass will have settled to its value today,   and
the axion field will oscillate around  the minimum with  a frequency
$\sim m_a$. 
The axions making up this field are  non-relativistic, because
their momenta $\lsim H_{QCD}\ll m_a$, where
\beq
H_{QCD} = \frac{1}{2t_{QCD}} = \frac{1.66\sqrt{g_*} T_{QCD}^2}{m_{pl}}
 \simeq 2\times 10^{-20} {\rm GeV } \frac{T_{QCD}^2}{(200 {\rm MeV})^2}
\label{HQCD}
\eeq
and  
$t_{QCD}\simeq  5$ km  is the age of the Universe, or
the horizon scale at the QCD  Phase Transition. This comoving scale
corresponds to $\simeq 0.1$ parsec today
(recall
the distance to the galactic centre is $\simeq 8$ kpc).
 
 The energy density in 
these  coherent oscillations redshifts 
 like matter, as $1/R(t)^3$ where $R(t)$ is the
scale factor of the Universe,
and   contributes to the
dark matter density today\cite{PDG}:
\beq
\Omega_a \sim  0.7  \times 
\left( \frac{f_{PQ}}{ 10^{12} ~{\rm GeV}}\right)^{7/6}
\left(\frac{a(t_{QCD})}{\pi f_{PQ}}\right)^{2}
~~~~~~~~~~ ~{\rm (coherent~ oscillations),}
\label{Oaosc}
\eeq
where $a(t_{QCD})$ is the value of the axion field
averaged over the Universe at the QCD phase transition.
Recall that $a/f_{\rm PQ}$   corresponds to the phase of
a complex scalar field, so could have any value
between $-\pi$ and $\pi$ in a causally connected volume.
In the case of interest here, when the PQ 
phase transition is  after inflation,
if one supposes  that all phases are equally
probable with linear measure, then  the  value
of the axion field, averaged over the Universe  is  
\beq
a(t_{QCD}) \simeq  \pi f_{\rm PQ}/\sqrt{3}~~.
\label{aQCD}
\eeq
Requiring that  the axions from topological
defects (\ref{Oastrings}) and the condensate  (\ref{Oaosc})
not over-contribute to   $\Omega_{CDM} \lsim 0.27$ 
 gives  $f_{\rm PQ} \lsim 10^{11}$ GeV \cite{strings}
in this case of the PQ transition after inflation.
On the other hand,
if the PQ transition is
before inflation, $a(t_{QCD})$ can be tuned to be
much smaller than $ f_{\rm PQ}$.  This is referred to as
the anthropic region of axion parameter space \cite{anthropic}, and
allows  values of $f_{\rm PQ}$  at the GUT scale.

Axions also have quartic interactions, as seen in eqn
(\ref{eqn3}). The coupling $m_a/f_{\rm PQ}$ is very small,
but for $a\sim f_{\rm PQ}$, the contributions to the
potential from the quartic and quadratic terms can
be comparable. We will neglect the quartic terms, because
we are interested in axion evolution after the QCD, so
the quartic term is suppressed by $[R(t_{QCD})/R(t)]^3$
with respect to the quadratic term (where $R(t)$ is the scale
factor, see eqn (\ref{metrique})).

We review in section \ref{ssec:TmnGR}  some results 
of density fluctuation growth
in a Universe whose CDM is axions.

Galactic halos made of BE condensed scalars,
of diverse masses and self-interaction strengths
(but not axions),
have been studied in \cite{RDS}, who confirm
the presence of vortices.
Galaxy formation  with axion-CDM has recently been studied by 
Banik and Sikivie \cite{BS13}.

\subsection{Bose-Einstein condensation}
\label{sec:BEC}

This section discusses what is a BE condensate, and some
approaches to 
calculating  how to get there.

The notion of a BE condensate, in equilibrium,  is familiar from the
statistical mechanics of particles:  in 
a thermal  bath with a sufficiently  large conserved charge
density,     the free energy 
is minimised if  charge-carrying bosons migrate to
the zero-momentum  state.

Equilibrium BE condensation is also a familiar notion for scalar
fields in cosmology. The classic papers of Kapusta
\cite{KapustaBEC} and Haber and Weldon \cite{HW},
evaluate the partition function for an 
interacting complex scalar field $\Phi$  at finite
temperature, in the presence of  a net charge density.
They  show that the chemical potential $\mu$  associated to
the conserved charge contributes a negative mass-squared 
to the  effective potential. So a sufficient charge density
can drive a phase transition,  to a  scalar
vacuum expectation value, which carries the excess charge not stored in the
equilibrium bath of particles: 
\beq
\label{becqft}
n_Q =
 m \langle \Phi \rangle ^2 + 
 \int\frac{d^3k}{(2\pi)^3} \left(
\frac{1}{e^{(E-\mu)/T} -1} - 
\frac{1}{e^{(E+\mu)/T} -1}
\right)
\eeq
where $n_Q$ is the charge density of the plasma.

If these equilibrium estimates are applied to
axions, after the QCD phase transition,
it is clear that  an axion number density $n_a \simeq m_a f_{\rm PQ}^2 \gg T^3$,
{\it if } in thermal equilibrium, 
must be in a BE  condensate. However,
axions are not thermally produced, and interact very
feebly.  ESTY  \cite{ESTY} propose
that they  ``thermalise gravitationally''. However
 \cite{ESTY,SY}  do not show that
the axion distribution approaches an equilibrium
distribution, or a migration of axion modes towards the infrared.

The particle and scalar field descriptions
of BE condensation  make contact in coherent
state notation \cite{I+Z}, 
in second quantised field theory, where  a coherent state
is denfined so that the expectation value of the field operator
gives the classical field (see eqn  (\ref{coherentstate})).
This illustrates the
observation of  Bogoliubov \cite{Bog}, that 
BE condensation  in non-relativistic
systems   can be described as  a phase transition.
 In the coherent
state perspective, the above two descriptions of 
BE condensation have two features:
\ben
\item a classical field is born from a state containing
particles. This requires $\hbar$, because $\hbar$ should be distributed differently in the Lagrangian to obtain particles or fields in the
classical limit \cite{hbar} (the mass has dimension length$^{-1}$
for fields). 
\item the particles move to a homogeneous and isotropic 
configuration where they are in their 
lowest energy state
\een

It is unclear to the authors which 
features  define a BE condensate, or, more precisely, what are
the   characteristics required of the axion dark matter
to allow caustic formation 
 as envisaged by  Sikivie and collaborators.
Is it coherence --- that is, a classical scalar field? 
Or is it a large population in the zero momentum lowest energy state?

If the crucial feature is coherence, then
 any (non-relativistic)
classical field would be a BE condensate. For instance, 
the axion field made via the misalignment mechanism, 
which can be decomposed on fourier modes, 
could   correspond to a superposition of BE condensates (one for
each three-momentum)\footnote{ In the approximation of this paper
where we neglect quartic axion  self-interactions, these
BE condensates have only gravitational  interactions
 with each other.}. This would be consistent with the detection
 of BE condensation in alkali gases \cite{Leggett}, demonstrated by
coherent collective behaviour of the atoms ( the BE condensate 
is allowed velocity). Maybe the two simple equilibrium  examples
of BE condensates, introduced above, are homogeneous and
isotropic because  equilibrium  is homogeneous and isotropic. 
 If the classical axion field is by nature a BE condensate,
then  the ``gravitational thermalisation'' of
misalignment axions is unneccessary and this paper is beside the point.

Experimentally,  BE condensation  occurs far
from  equilibrium \cite{Leggett,stoof,livre96}. 
Theoretically,   a  Closed-Time-Path \cite{C+H} implementation of
 the 2 Particle Irreducible effective action \cite{CJT}
(see {\it e.g.}
the chapter on this subject in  \cite{C+H}), 
 allows to compute the  out-of-equilibrum
generation of a BE condensate. 
The 2PI effective action is a function of  both the 
classical field and of  the  two point function (and the two point function in
closed time path  represents 
the number density and propagator). 
Analytic calculations  have been performed in
self-interacting scalar field theories \cite{BergesSerreau}, 
and show \cite{latticephi4} that at NLO,
an overpopulation of low momentum modes in
the number density can
institute an inverse cascade towards the infrared,
without first establishing an equilibrium distribution.
Recall, however, that a high density of low momentum
modes is not a classical field (or a BE condensate);
it lacks the required coherence.

 In summary,  we focus on the gravitational interactions
of the  misalignment axions, which are already
a classical field.
We look for dissipation  in these interactions,
 because this increases entropy.
In 2PI formalism,
thermalisation   does not occur at leading order
in the coupling in  $\phi^4$ 
models; extrapolating naively, this suggests that
gravitational thermalisation, or entropy generation, 
does not occur at leading order in $G_N$.
Instead of including ${\cal O}(G^2_N)$, we look
for dissipation/thermalisation at order $ G_N |\vec{p}|^2/m_a^2$.
{ This could be  relevant to BE condensation, if the
axion field produced by the misalignment
mechanism  is not already a BE condensate. 
(This assumption  is consistent  with \cite{SY}.
References \cite{SY1,ESTY} envisage, that
in this case,  the ``gravitational thermalisation'' of axions
will drive them to a BE condensate.)}

\subsection{Axion Bose-Einstein condensation  in cosmology}
\label{sec:bose}

This project was motivated by  the scenario envisaged in
  \cite{SY1,ESTY}, where    
 axions form a  Bose-Enstein condensate
in the early Universe at photon temperatures $T \sim $ keV,
due to gravitational scattering among the axions.
The gravitational interaction rate of  \cite{SY1,ESTY}
was confirmed by Saikawa and Yamaguchi (SY) \cite{SY},
who calculate in Quantum Field Theory, the  rate of
change of the axion number operator $-i [\hat{H},\hat{n}(k)]$.
SY  describe the axions  as a coherent state (see
eqn  (\ref{coherentstate})) in Minkowski space-time, 
interacting via Newtonian gravity. 
This significant calculation has various curious features: 
the  Newtonian analysis  is applied in the early Universe,
without a distinction between the homogeneous energy
density which drives expansion, 
and the 
fluctuations. 
Also,  intuition and the
equations of linear structure growth
say that gravity grows inhomogeneities,
which  appears naively  at odds with gravitational
interactions driving axions to  Bose-Einstein condense in 
the zero mode.  Another curious feature is 
that, although gravity should be universal,
 the axions are found to ``gravitationally thermalise''
with themselves, but not with other particles.  
We discuss the  interpretation of our estimates and
these earlier calculations in section \ref{sec:disc}.

Finally, we  raise one more confusing issue.
 A BE condensate  in statistical mechanics
is  a large number of particles 
in a $\delta$-function  at zero kinetic energy. In the coherent state
notation of eqn (\ref{coherentstate}), these  particles
make up the first term of  eqn(\ref{becqft}).
However,  in cosmology, it is unclear  how narrow is 
the energy range for  the axions making up the ``zero mode'',
or BE condensate. At the QCD phase transition,  
the  axions  of mass $m_a$ and momentum $H_{QCD}$, 
have  kinetic energy  $E_K =H_{QCD}^2/2m_a \ll H_{QCD}$.  
During radiation domination, the ratio
\beq \frac{E_K}{H} \simeq \frac{H_{QCD}}{2 m_a} \ll 1 \label{E/H} \eeq
remains constant; between matter-radiation equality and today, it increases by a factor $\sqrt{T_{eq}/T_0}$, but does not attain one. Therefore, the Heisenberg
uncertainty principle  could imply that  the age of the Universe 
is not long enough to distinguish that the axions are not in the zero mode.
Does this imply that they  are in a BE condensate? 
Notice that their three-momentum can be distinguished from
zero, so if the condensate was defined as the zero-momentum
state, then  the axions are not in it.


\section{Estimating axion viscosity}
\label{sec:Tmn}

This section aims to address  whether
classical gravitational interactions among axions
can dissipate fluctuations and generate entropy. The Appendix 
suggests that  in a  homogeneous and
isotropic Universe, 
the answer is ``no'' (gravity 
merely redshifts  the axion momenta,
in a Friedman-Robertson-Walker
Universe). So we study a 
cosmological scenario with  density
fluctuations, whose  gravitational effects 
can be treated in the linear approximation. 
The question would be more difficult, in the case where  gravity
is non-linear.

We take  an initial condensate made of
axions with comoving momenta  of order $H_{QCD}$. 
We compute their stress-energy tensor 
in an almost  homogeneous and isotropic
Universe, including scalar metric perturbations
in Newtonian  gauge \cite{mukhanov,MB}.
 In particular, the metric perturbations will give
spatial off-diagonal components  $T^i_j$, for $i \neq j$. 
We 
equate the term involving the gravitational potential
with the off-diagonal  $T^i_j$
 of an imperfect fluid in a homogeneous
and isotropic Universe. In an imperfect fluid,
the $T^i_j$ elements, for $i \neq j$,  are proportional
to the viscosity, which damps short-scale fluctuations. 
 The only interactions of
the axions are gravitational, so implicitly,
the ``bath'' responsible for dissipation in
the fluid contains metric and density fluctuations.
Equation (\ref{dissipation}) is an estimate of 
the damping scale of  density fluctuations,
 due to the gravitational self-interactions of axions.

\subsection{Axion initial conditions after the QCD phase transition}
\label{ssec:ic}

We take our initial conditions a few Hubble times after
the QCD phase transition, when the axion mass is settled to
its value today. 
We focus on the classical axion field produced
by misalignment, this does  not include
the axions from strings.  Classical field
means  in principle the variable in the 1PI
action, and in practise the expectation value
of the field operator in the ground state. It
can be expressed as a coherent state, see eqn
(\ref{coherentstate}). 
We suppose that at the QCD phase transition,
 the axion   field was
approximately constant within a horizon volume, 
and randomly distributed between $-\pi$ and
$\pi$ from one horizon volume to another. So  the initial
axion field oscillates rapidly in time with frequency $m_a$,
and more slowly in space with co-moving momentum $\sim H_{QCD}$.
It can be expanded  on   Fourier modes  
of    a homogeneous and isotropic Universe:  
\beq
a(\vec{x},t) 
= 
 \frac{ 1}{\sqrt{2mVR^{3}(t)} } 
\sum_p  \left[ \widetilde{a}(\vec{p},t) 
\exp\{  i(\vec{p}\cdot \vec{x} - \omega t)\}
+ \widetilde{a}^*(\vec{p},t) 
\exp\{ - i(\vec{p}\cdot \vec{x} - \omega t)\} \right]
\label{phift}
\eeq
where  $\vec{p}$ is the comoving three-momentum,
and the field is normalised in a comoving box of
volume $V$. Recall that $a(\vec{x},t) $ has 
mass dimension one, so  
the 
  $\widetilde{a}(\vec{p},t)$
are dimensionless, and 
 $|\widetilde{a}(\vec{p},t)|^2/2$
is the number of axions of momentum $\vec{p}$
in the volume $V$ (see eqn (\ref{na})).
The fast time dependence
$e^{-i\omega t}$,  which  we approximate as
$e^{-im t}$ can be averaged \cite{aCDM,linearaxions,ratra}
 on the longer  evolution timescale 
of the spatial variations\footnote{It can
be removed more elegantly by studying
the non-relativistic field \cite{NambuSasaki}.}.  $\widetilde{a}(\vec{p},t)$
can  evolve in time on this longer timescale.
The  axion is a real field, so 
the Fourier
coefficients satisfy   $\widetilde{a}(\vec{p},t)=
\widetilde{a}^*(-\vec{p},t)$.
Fourier transforms are performed  in a comoving box $V=L^3$,
and defined to be ``dimensionless'' to simplify
dimensional analysis;
we write
$$
\frac{d^3x}{V} ~~~, ~ {\rm and} ~~~ 
\sum_p = V \int \frac{d^3p}{(2\pi)^3} ~~~.
$$

Notice that the density fluctuations are of order one
on the comoving scale $H^{-1}_{QCD}$: the field can be zero in one
horizon volume, and $\pi$ in the next.
We are  interested in comoving distances
longer than a few $\times H^{-1}_{QCD}$, so
we  take  $\widetilde{a}(\vec{p},t) \to 0$ for
$|p_i| \gg H_{QCD}$ and expect 
 that  it is flat for    $|p_i| \lsim  H_{QCD}$,
because  the Fourier transform
of a random distribution in $x$ is
a constant \footnote{This means that fluctuations
get smaller on larger distances $L$: 
$\int_{L^3} \delta \rho \sim \overline{\rho}/\sqrt{L^3H^{3}_{QCD}}$.}.

Equation (\ref{phift}) appears 
different from the usual treatment of axion CDM, where (most of) the axions
are taken to be in the zero-momentum mode(see {\it e.g.}
 \cite{HwangNoh}):
\beq
a(\vec{x},t) 
= 
 \frac{ 1}{\sqrt{2mVR^{3}(t)} } 
\left(  \widetilde{a}_0(t_{QCD})  \cos (mt) + 
\sum_p [\delta \widetilde{a}(\vec{p},t) e^{i(\vec{p}\cdot\vec{x} -mt)}
+ \delta \widetilde{a}^*(\vec{p},t) e^{-i(\vec{p}\cdot\vec{x} -mt)}]
\right)
\label{ahomft}
\eeq
where $ \widetilde{a}_0(t_{QCD})/\sqrt{2mV} =
{a}(t_{QCD})$ is the averaged-over-the-Universe
 value of the field  at the QCD phase transition, and the
small fluctuations considered are on  large scale structure
scales. 
As discussed in \cite{BGBL},  the difference between these two forms for
the growth of linear density perturbations is negligible:
the kinetic energy density  in the horizon-scale fluctuations is
negligible (compared  to the potential and the
time oscillations), and  on structure formation scales,
 ${\langle a(\vec{x},t)}  \rangle (t_{QCD}) \simeq \pi f_{\rm PQ}/\sqrt{3}$.

\subsection{The stress-energy tensor with perturbed metric}
\label{ssec:TmnGR}

This section   reviews the stress-energy tensor
and equations of motion for scalar perturbations in 
an  almost homogeneous and isotropic
Universe.

The metric in Newtonian gauge  can be written
\bea
ds^2 & = & (1 + 2\psi)dt^2 - R^2(t) (1 - 2 \phi) \delta_{ij} dx^i dx^j 
\label{metrique}
\eea
where $\phi \simeq \psi$ will be the
Newtonian potential inside the horizon, 
and we take  the scale factor $R(t)$ dimensionless
and  equal to 1  at the QCD phase transition.

The
stress-energy tensor 
for a homogeneous and isotropic Universe is
$T^{\a}_{\b} = {\rm diag} (\overline{\rho}, -\overline{P},
-\overline{P},-\overline{P})$,  
where $\overline{\rho}$ and $\overline{P}$ are the 
(homogeneous and isotropic) 
 energy density and pressure. 
In the presence of  scalar fluctuations, 
$T^{\a}_{\b}$  can  be described with
four additional parameters, written  in Fourier space as
 \cite{MB}
\bea
\overline{\rho}(t) \to \overline{\rho}(t) + \delta \tilde{\rho}(\vec{k},t)
~~~,~~~
\overline{P}(t) \to \overline{P}(t) + \tilde{\delta P}(\vec{k},t)
\nonumber \\
ik_j \delta T^0_j = (\overline{\rho} +\overline{P}) \theta (\vec{k},t)
~~~,~~~
(\hat{k}_i \hat{k}_j - \frac{1}{3}\delta_{ij})  \delta T^i_j = 
-( \overline{\rho} +\overline{P})\sigma  (\vec{k},t)
\label{fluide}
\eea
where $\theta$ parametrises a fluid velocity, and
$\sigma$ is the anisotropic stress.

For a massive non-interacting  real  scalar field,
such as the axion, the stress-energy tensor has the form 
\beq
T^\mu_\nu= a^{;\mu} a_{;\nu} -
   \frac{1}{2} \left(  a^{;\a}
a_{;\a} - m^2 a^2 \right) \delta^\mu_\nu~~~.
\label{TmnGR}
\eeq
Equating (\ref{TmnGR}) and  (\ref{fluide}) allows to
determine the  density fluctuations and other fluid
parameters of the classical axion field.

For
axions of the form given in eqn (\ref{phift}), $\overline{\rho}_a(t)$
 is the  $\vec{p} = 0$ Fourier mode of the density 
$\widetilde{\rho}(\vec{p},t)$:
\bea
 \overline{\rho}_a(t)& =&
\int_V \frac{d^3 x}{V} T^0_0(\vec{x},t) \nonumber\\
&=&
\frac{m^2}{ [R(t) ]^3}
 \sum_q   \frac{|\widetilde{a}(\vec{q},t)|^2} {mV}  \left( 
1+ \frac{q^2}{m^2R(t)^2} \right)  +...
\label{rhokk3}
\eea
where the $+...$ contains subdominant terms  involving
$H$, $\phi$ and $\psi$, and the $q^2/m^2$ term will
be neglected.
The (volume-averaged)  number density of axions
in the classical field  can
similarly be expressed as
\bea
n_a(t)& =&
\frac{m }{ [R(t) ]^3}
 \sum_q   
 \frac{|\widetilde{a}(\vec{q},t)|^2} {mV}
  +...   ~~~.
\label{na}
\eea
If $n_a(t_{QCD}) \simeq m\pi^2 f_{\rm PQ}^2/3$,
and  $\widetilde{a}(\vec{q},t)$ is approximately
constant for $|\vec{q}|\lsim \sqrt{3} H_{QCD}$, then
 $|\widetilde{a}(\vec{q},t)|^2 \simeq $   
$\frac{\pi^4m_\pi f_\pi f_{\rm PQ}}{H^3_{QCD} } $.

The 
 Fourier transform of the
density fluctuations in the classical axion field is
\bea
\delta \widetilde{\rho}_a(\vec{k},t)
&=& 
\int \frac{d^3 x}{V} e^{-i \vec{k} \cdot \vec{x}}
[ \rho_a(\vec{x},t) - \overline{\rho}_a(t)]
\nonumber\\
&=&
\frac{m^2}{ mV [R(t) ]^3}
 \sum_q 
 \widetilde{a}(\vec{q}+\vec{k}/2,t) \widetilde{a}^*
(\vec{q}-\vec{k}/2,t)   
~~~~k \neq 0
\label{drho}
\eea
where  we have dropped terms proportional to  
$H$, $\phi$ and $\psi$, and $\{q^2, k^2\}/m^2$.
Notice that this formula is different (less intuitive) 
from the   case usually studied in structure
formation, where most axions are in a  zero-momentum
condensate. If most axions are in  the zero mode, the density
fluctuations  on scale
$k^{-1}$  are linear in the field fluctuations,
 $\delta\widetilde{\rho}(\vec{k},t) \sim  \widetilde{a}_0 \delta \widetilde{a}(\vec{k},t)$
so are made up of axions of momentum $k$ (in the coherent
state formalism of eqn (\ref{coherentstate})).

The dynamics is controlled by  
 $ T^{\a\b}_{~~~;\b} =0$, and by
Einstein's Equations 
$ G_{\a\b}  = 8 \pi G_N T_{\a\b}$. 
In the absence of perturbations,
these give the Hubble expansion rate
\beq 
\left(\frac{\dot{R}}{R(t)}\right)^2
\equiv
H^2(t) =  \frac{8 \pi G_N}{3} (\overline{\rho}_a(t) +
\overline{\rho}_{strings}(t) + \overline{\rho}_{rad}(t)) ~~~.
\label{EqnH}
\eeq
where  $\overline{\rho}_a(t) + \overline{\rho}_{strings}(t) +
\overline{\rho}_{rad}(t)$ is 
the Universe-averaged density in the  axion field,
in the axions from strings,  and in radiation.

The equations for  the scalar metric  and
density fluctuations can be found in \cite{mukhanov,MB,K+S}.
For the stress-energy tensor of the axion field,
eqn (\ref{fluide}), the condition  $ T^{\a\b}_{~~~;\b} =0$,
gives the scalar equation of motion in the
perturbed Universe.
Whereas for  a perturbed fluid,
eqn (\ref{TmnGR}),   $ T^{\a\b}_{~~~;\b} =0$,
gives two equations for the four parameters.   If the  speed of
sound $c^2_s = \delta P/\delta \rho$ can be calculated,
and $\sigma$ neglected, then  to determine
the dynamics of the fluid (like those
of the field),  requires only one additional
equation from  Einsteins Equations. 
 We will be interested in fluctuations inside the horizon, so neglecting
terms of order $\partial_t\phi,  H(t)$,
the Einsteins Equations give, in Fourier space,  the Poisson equation for 
$\widetilde{\phi}$:
\beq
-\frac{|\vec{p}|^2}{R^2(t)} \widetilde{\phi}(\vec{p},t) \simeq 
4 \pi G_N \delta \widetilde{\rho}(\vec{p},t)
\label{newton}
\eeq
where $ \delta \widetilde{\rho}(\vec{p},t)$ is the Fourier
transform of the density fluctuation (in radiation and
axions). Notice that this can be interpreted as the
potential due to single graviton exchange \cite{PS125},
which we will allude to in the discussion.

Combining the various equations gives the well-known
equation \cite{linearaxions,ratra,HwangNoh}  for  the evolution of adiabatic 
scalar  
density fluctuations
$\delta\equiv \delta\rho(\vec{k},t)/\overline{\rho}(t)$:
\beq
\ddot{\delta} + 2H\dot{\delta} -4\pi G_N \overline{\rho} \delta
+c_s^2 \frac{k^2}{R^2(t)} \delta = 0
\label{ddd}
\eeq
The equations for isocurvature fluctuations are different
\cite{BE86}, but they share with this
equation the property that density fluctuations
are frozen within the horizon during radiation
domination\cite{BE86}, and can grow during
matter domination. 
For  a homogeneous axion field with small
fluctuations, as would arise if the PQ phase
transition was before inflation  (see eqn (\ref{ahomft})),
the equation (\ref{ddd})  is  elegantly obtained in \cite{HwangNoh}.
In this case,  it is shown that $\sigma = 0$
and $c_s^2 \simeq  k^2/(4m^2 R^2(t))$.
The (physical) axion Jeans length is therefore 
\beq
\lambda_J (t) \simeq  \frac{2\pi}{[16 \pi G_N \overline{\rho}(t) m^2]^{1/4}}
\sim \frac{6}{\sqrt{H(t) m}}~~~;
\label{jeans}
\eeq
on shorter distances, the fluctuations oscillate due
to axion pressure, on larger distances, they can grow in
a matter-dominated Universe.
It can be checked that
$\lambda_J \sim   \sqrt{{H_{QCD}}/{m}}  \times
H^{-1}_{QCD} R(t)$, suggesting that axions behave
like dust on 
the comoving distance of the QCD horizon.

A caveat is that $\sigma$ and $c_s$ may be different
for the axion field configuration  arising when the PQ transition
is after inflation (see eqn (\ref{phift})). However,
by dimensional analysis, $\sigma \lsim H_{QCD}^2/m^2$
and $\overline{P} \sim \delta P \lsim  \overline{\rho} H_{QCD}^2/m^2$,
so they naively appear insignificant to fluctuation
evolution on the comoving scale $H^{-1}_{QCD}$.

Recall that the fluctuations in the density
of the axion field on
the scale $H^{-1}_{QCD}$ are of ${\cal O}(1)$.
After matter-radiation
equality, these short-distance isocurvature fluctuations
can grow and promptly decouple from the Hubble flow,
to form gravitationally bound axion configurations
called ``miniclusters'' \cite{mini}. The
miniclusters can further cool and contract
due to gravitational interactions \cite{SeidelS}.
The position-space perspective on these
 ${\cal O}(1)$ inhomogeneities is instructive. 
One can estimate that an axion (particle?) with comoving momentum
 $H_{QCD}$ cannot escape from a fluctuation of
comoving size  $H^{-1}_{QCD}$ prior to matter-radiation
equality.  That is, the fluctuations
are not damped by free-streaming.  If an axion
BE condensate should be approximately homogeneous,
then it is unclear to
the authors how the axions making up 
 ${\cal O}(1)$ density fluctuations on scales
 $H^{-1}_{QCD}$ can migrate to the zero-momentum mode,
because they do not seem to move fast enough to
homogenise in position space. 

\subsection{Anisotropic stress }
\label{ssec:anis}

The previous section showed that  the small $T^i_j$ elements
of the  stress energy tensor of the classical
axion field  where unimportant for
fluctuation growth. This section calculates these
off-diagonal elements, with the aim of identifying
in them some gravitational dissipation.

The  off-diagonal spatial elements
$T^i_j$ are interesting  for two reasons:  
they are gauge invariant, and  in the fluid
approximation,  they are proportional to
the viscosity. Viscosity damps fluctuations on small
scales \cite{Weinberg},  so we hope that an estimate of
the viscosity  will give some notion of gravity's ability
to generate entropy. 
The first step is to compute $T^i_j(\vec{x},t)$, for
$i\neq j$:
\bea
T^i_{~j} (\vec{x},t)  =
-\frac{(1 +2\phi)}{R^2(t)} \partial_{i} a
 \partial_{j} a
\label{aniP}
\eea
so in Fourier space:
\bea
T^i_{~j} (\vec{k},t) & =&  
-\frac{1}{mVR^5(t)}
 \left[  \sum_{q}
(q+k/2)_i   (q -k/2)_j  \widetilde{a}(\vec{q} + \vec{k}/2,t)  
\widetilde{a}^* (\vec{q} - \vec{k}/2,t) \right.
\nonumber\\
&& ~~\left.+ 2  \sum_{p,q} 
 ( q + k/2)_i (q +p -k/2)_j \widetilde{\phi} (\vec{p},t) 
 \widetilde{a}(\vec{q} + \vec{k}/2,t)  
\widetilde{a}^* (\vec{q}+ \vec{p} - \vec{k}/2,t) 
\right]~~~.
\label{aniP2}
\eea
We drop the  first  term (not involving
the metric fluctuation)  in this expression,
which arises because the condensate of axions
with finite momentum is not perfectly homogeneous
and isotropic.  This term is
naively of order the  axion  pressure $\sim H_{QCD}^2/(R^2(t) m_a^2)
\times \overline{\rho}_a(t)$, which we also
neglect. This first term   in principle contributes 
to distinguishing  
$\phi$ from $ \psi$   (see the metric of eqn(\ref{metrique}))
in  the perturbed Eintein Equations :
\beq
k_i k_j (\widetilde{\phi}(\vec{k},t) - \widetilde{\psi}(\vec{k},t))  = 
 12 \pi G_N  T^i_j ~~~~~(i \neq j)~~~.
\label{whatabout}
\eeq
However, we neglect this effect and take 
$\phi = \psi$,  
 because  the axions constitute initially
a tiny fraction of the
energy density,  and this contribution
to $T^i_j$ decreases  as $1/R(t)$ compared to
the total density (second order
radiation perturbations could be more significant).
The second term of eqn (\ref{aniP2}), which contains
the gravitational potential of density perturbations,
is the piece from which we wish to extract
axion viscosity.

\subsection{Matching to an imperfect fluid}
\label{ssec:TmnW}

We aim to  obtain a viscosity coefficient for our axion fluid,
despite that we do a  classical field analysis with coherent
initial conditions. We need fluctuations and dissipation, 
and since we approximate the axions to have only gravitational
interactions, these must involve gravity. 
We therefore  map the stress tensor of
the Universe with metric fluctuations,
 onto the   imperfect fluid  stress tensor of a homogeneous
and isotropic Universe. One can imagine that
 the density/metric fluctuations generate the viscosity.

The stress-energy tensor for an imperfect fluid in a
homogeneous and isotropic expanding Universe is given in
\cite{Weinberg}. For $i\neq j$:
\bea
T^i_j(\vec{x},t) &=& -\eta(t) (\partial_jU^i(\vec{x},t) 
+\partial^iU_j(\vec{x},t))
\label{W1}
\eea
where   $U^\a$ is the fluid four-velocity, which
Weinberg defines from 
the conserved number current  $N^\a = n U^\a$ ($n^2 =  N_\a N^\a$). 
Since axions are a real field, it is convenient
to use an alternative definition, so we define
 $U^\a$ from the energy flux: $T^0_i = \rho U^0 U_i$.
This has the added interest of
 giving a $k_ik_j$ term in eqn(\ref{W2}). 
As discussed with care in Weinberg's paper
\cite{Weinberg},
it is important to use a self-consistent
formalism, so we anticipate that our
estimate will not have the correct
constant factors. We hope that the
dependence on physical parameters will
nonetheless be correct.
For non-relativistic axions, eqn(\ref{TmnGR})  in
a homogeneous and isotropic Universe gives
\beq
U_0 U^i (\vec{x},t)  \simeq \frac{-1 }{R^2(t)\overline{\rho}(t)}
\partial_t a(\vec{x},t) \partial_i a(\vec{x},t) 
\eeq
which gives 
\bea
T^i_j(\vec{k},t) &=&  -\frac{\eta(t)}{mV R^5(t) n_a(t)} 
  \sum_{q}  \left[q_i k_j  + k_i q_j -k_i k_j  \right] 
\widetilde{a}(\vec{q} + \vec{k}/2,t)  
\widetilde{a}^* (\vec{q} - \vec{k}/2,t) 
\label{W2}
\eea
with $n_a(t)$ from eqn (\ref{na}).

Equating the coefficients of $\hat{k}_i\hat{k}_j$ in  
eqns (\ref{W2})  and  the second line of eqn(\ref{aniP2}),
 gives
\beq
\frac{\eta(t)}{ n_a(t)} \sim 
- 2 \pi G_N  \sum_p
\frac{ \delta \widetilde{\rho}( p,t) R^2(t)}{ |\vec{p}|^2}
\label{eta}
\eeq 
where we suppose the  $p$ in the sum  on $q$  of $|\widetilde{a}|^2$
makes little  difference.

This estimate used a description of imperfect fluids \cite{Weinberg}
 which can suffer from 
non-causal information propagation. Such
difficulties are avoided  with 
the causal thermodynamics of \cite{Maartens:1996vi}, 
which adds approximately a factor $ (1 + H / \Gamma_g)$ 
to the right side of (\ref{eta}), where $\Gamma_g \sim 8\pi G \rho_a m_a R(t)^2/H_{QCD}^2$ is 
the  gravitational interaction rate of axions
(see eqn \ref{finzzz}). 
The correction factor exceeds 2 for $T > 2 \mbox{keV}$ (for $f_{PQ} \sim 10^{12} \mbox{GeV}$ fixed)
and grows linearly with $T$.
 However,  we neglect this effect, because
it never allows the time or length scale of
dissipation to reach the horizon, and because
axions gravitationally thermalise after $T \sim$
keV in the scenario of Sikivie and collaborators. 

There are two simple limits for the estimate of eqn (\ref{eta}). First, if
the dominant  density fluctuations are the no-scale
adiabatic fluctuations in the radiation,
then the sum
is infrared divergent and dominated by horizon-scale
fluctuations\footnote{If $k_H$ is the comoving  scale
at the horizon,  we can write
$\delta \widetilde{\rho}(k,t)/\overline{\rho}(t) =  A(k_H/k)^{3/2}$, to obtain 
a no-scale power spectrum such that $V\int d^3k P(k) =$ 
$ 4 \pi Vk^3_H |A|^2 \ln k_{max}/k_{min}$. Then
$V \int_{k_H} d^3p \delta(p,t)/p^2= 4 \pi Vk_H A \sim
\delta \widetilde{\rho}(k_H,t)/(\overline{\rho}(t) k_H^2)
$ }:
\beq
\left|\frac{\eta(t)}{ n_a(t)}\right| \sim 
2 \pi G_N  
\left|\frac{  \delta \widetilde{\rho}(H(t)\frac{T_{QCD}}{T},t)}  {H ^2(t)}\right|
\simeq
\frac{3}{4}\left|\frac{ \delta \widetilde{\rho}(H(t)\frac{T_{QCD}}{T},t)}{\rho(t)}\right|
\label{etarad}
\eeq 
Before and during linear fluctuation growth,
this gives $\eta(t) < n_a(t)$.

The second case is when the dominant density fluctuations
are the axion inhomogeneities on the co-moving scale
$H_{QCD}^{-1}$. Then the sum 
$d^3 p \delta \widetilde{\rho}_a/|\vec{p}|^2$
in eqn (\ref{eta})
is dominated by $p \sim H_{QCD}$, giving
\bea
\left|\frac{\eta(t)}{ n_a(t)}\right| \sim 
2 \pi G_N  
\left|\frac{ \delta \widetilde{\rho}(H_{QCD},t) R^2(t)}{H_{QCD}^2}\right|
\lsim \frac{8\pi G_N  \rho _a(t_{QCD}) }{3 H_{QCD}^2} 
\left( \frac{T}{T_{QCD}} \right)
&=&\frac{ \rho _a(t_{QCD}) }{\rho_{rad}(t_{QCD})}
\left( \frac{T}{T_{QCD}} \right) 
\nonumber\\
&=& \frac{ T_{eq} T}{T^2_{QCD}} 
\label{etaax}
\eea 
where
 $T_{eq}$ is the photon temperature at
matter-radiation equality.

Weinberg gives \cite{Weinberg}  that modes of 
comoving   wavenumber $\vec{p}$ decay
at a rate
\beq
\Gamma \sim \frac{\eta(t) |\vec{p}|^2}{R^2(t)\overline{\rho}(t)}
\eeq
So  the physical distance
 $\ell_{damp} (t)$ on which fluctuations could disappear
grows as the square  root of the time available. This makes
intuitive sense when fluctuations are damped by
particles random-walking out\footnote{ Recall, however,
that this picture corresponds to perturbation
theory in the mean free path $\sim 1/\langle \sigma n\rangle$.
So weakly interacting particles diffuse more
easily out of perturbations, and the
interpretation is unclear  for particles whose
mean free path is the size of the perturbation.}.
  In the axion case
studied here,  
in the lifetime of the Universe $\sim 1/H$,
fluctuations on physical distances less than
 $\ell_{damp}$ could dissipate, where
\beq
\ell^2_{damp}(t=1/H)  \sim  
\frac{1}{H(t) m_a}
 \frac{\eta(t)}{n_a(t)}\frac{\rho_a(t)}{\rho(t)}
\label{dissipation}
\eeq 
It is clear that these estimates give a 
damping distance  much shorter  
than the comoving scale $H_{QCD}$.
The Jeans distance for axions is $1/\sqrt{H(t)m}$ \cite{linearaxions};
at shorter distances, density fluctuations in axions oscillate
due to pressure,
and at larger distances the fluctuations can grow 
(during matter domination).
 It is reassuring  that the
damping distance due to viscosity  (\ref{dissipation}) is shorter than 
the Jeans length.


\section{Discussion and comparison to previous results}
\label{sec:disc}

The  estimated 
damping scale   (\ref{dissipation})
 for axion density  fluctuations
prior to the period of non-linear structure formation, 
is always  shorter than the QCD horizon scale $T_{QCD}/(TH_{QCD})$. 
It does not  confirm that   ``gravitational thermalisation''
erases axion fluctuations on the QCD horizon scale 
at $T \sim$ keV. We first comment on our estimate,
then compare to the calculation of Saikawa and Yamaguchi
\cite{SY}.

\subsection{Our  estimate}

A first simplifying approximation made  in this paper, is
that we only work to linear order in $G_N$.  This could appear
curious compared to thermalisation rates 
associated to  Boltzmann Equations, where the rates
are proportional to couplings-squared. 
But classical fields, expressed as coherent
states, correspond to  the coherent superposition
of amplitudes, so classical gravitational effects
appear at linear order in $G_N$ (as is well known).

We focus on  the axion energy density, and
fluctuations therein, rather than on the axion field.
It is clear that in general, the field carries
more information than the energy density, since it
allows to compute a wider variety of correlation functions. 
However, at the classical level used in this paper,
the equations of motion for both the axion field and
the density fluctuations are obtained from
$T^{\mu \nu}_{~;\nu} = 0$ and the Poisson Equation 
(\ref{newton}), which suggests that it is merely
two different parametrisations of the same physics
\footnote{Whether this formulations are equivalent is
important, because 
 BE condensation
corresponds to suppressing  the field fluctuations.
One can wonder if  gravitational
interactions could homogenize the field configuration
without changing the stress-energy tensor. 
The linearised Einsteins Equations might  suggest not: 
the stress-energy fluctuations induce  the Newtonian potential. 
}. 
The equations for the density
fluctuations have the advantage that
they are linear and can be solved. 
They say that gravity grows inhomogeneities
in axions. Whereas the equations
for the field are  non-linear; 
 a gravitational interaction
rate for axions can be calculated
without solving the equations, 
but that does not say what gravity does with the axions. 
 So we do not disagree with the gravitational
interaction rate of axions obtained by
\cite{ESTY,SY} (we can reproduce it, see
eqn (\ref{finzzz})); however, we disagree 
with its interpretation as a thermalisation
rate. We suspect that it is the rate associated
with the gravitational growth of density fluctuations (which
is compensated by the expansion of the universe during radiation
domination).

We supposed that BE condensation 
requires dissipation, and  furthermore, that
leading order  solutions  of classical
equations of motion do not exhibit dissipation
or thermalisation. This is a usual
perspective in  non-equilibrium field theory ---
to obtain dissipation from time-reversal invariant
equations requires summing over a bath
of fluctuations. We are unclear on how
to separate gravitational interactions in
the early Universe into a leading order
solution plus fluctuations that we can integrate.
 Therefore, we hesitate to discuss  
a ``gravitational thermalisation'' rate, because
its definition seems to require this separation
of gravitational interactions into ``leading order''
and ``dissipative''. 
It may be unwise to identify the homogeneous and isotropic
component of the Universe as the leading order
solution, and the fluctuations as the bath,
because density fluctuation
growth is an important part of the classical solution.
  However, the ${\cal O}(|\vec{p}|^2/m_a^2)$ terms
are usually neglected in these equations,
so we attempt to associate dissipation with them:
in the perturbed, expanding
Universe, the off-diagonal spatial elements
of the  stress-energy tensor are gauge invariant,
of  ${\cal O}(|\vec{p}|^2/m_a^2)$, 
and unimportant for fluctuation growth. 
We identify them with the off-diagonal
elements of the stress energy tensor of an
imperfect fluid. An imperfect fluid can
grow density fluctuations, but contains
dissipation, so we hope, by this
identification, to  be summing  over
the  gravitational fluctuations that
are not an  important part of the classical
solution.  Fortunately, the
damping scale we obtain is
irrelevantly short, so whether
this trick  is credible is of minor importance.

It can be useful to compare to classical thermalisation
studies in $\phi^4$ models \cite{aart}, using the 2PI action
 \cite{CJT} in closed time path formalism. 
In this formalism, the dynamical variables
are the classical field and
the two point function (which describes
the density of incoherent modes and the propagator).
Intuitively, one could anticipate that the
classical field could dissipate
by interacting with the bath of incoherent
fluctuations. In this paper, we neglected the
cold axion particles produced by strings, which could
be the bath thermalising the axion field (because 
at linear order in $G_N$, they should just constitute
 an additional contribution to  density fluctuations,
with which the fluctuations in the density of the axion
field   could interact).
Studies of thermalisation
in $\phi^4$ find that the incoherent
modes thermalise at NLO \cite{BergesSerreau}. So  perhaps it
would be interesting  to study the evolution of
axions from strings and misalignment using
the 2PI effective action, in the Closed Time Path
formalism  used by  Saikawa and Yamaguchi.

\subsection{Making contact with previous calculations}

We now address the differences between our estimate 
and the  calculation of Saikawa and Yamaguchi (SY) \cite{SY}.
{ We focus on this impressive analytic calculation,
because they introduce very clearly the used formalism 
and obtain the same result as \cite{ESTY}.  }
 SY calculate the time evolution
of the axion number operator,
using  a closed time path formalism
of Quantum Field Theory, 
in flat space-time with Newtonian gravity.
They evaluate $dn(\vec{q})/dt$ (which is the rate of change of 
the number density  of axions of momentum $\vec{q}$
due to gravitational interactions), 
in a coherent state representing highly
populated low-momentum axion states.
  This rate is interpreted
as an axion thermalisation rate,
and it is larger than $H$ for photon
temperatures $\lsim$ 1 keV.

An unimportant  difference  is that 
the redshifting due to Universe
expansion does not appear in the SY calculation. The
 gravitational effect of the
homogeneous and isotropic axion
density  is to drive expansion,
but since  SY  calculate 
with Newtonian gravity in a non-expanding space-time,
all the gravitational effects
of the  axions are included in 
the  ``thermalisation''
process. This is a relatively minor issue;
scale factors can be  judiciously
distributed in their formulae, and
within the horizon, density fluctuations
can be described by  Newtonian gravity. 
If  the density $\rho(x,t)$ in the SY formulae
is replaced by the density fluctuation
 $\delta \rho(x,t)$, then their
equations are consistent with the classical linearised
Einsteins Equations in Newtonian gauge.

An obvious difference  from our classical
discussion is that SY calculate in quantum field theory.
This seems also to be unimportant, because { using classical equations of motion} we
can obtain a similar
\footnote{From eqn (\ref{na}),  
$ |\widetilde{a}(\vec{q},t)|^2/ (\sum_p |\widetilde{a}(\vec{p}),t|^2)$  
is the fractional number density of 
axions of momentum $\vec{q}$. This equation can be obtained
from the equations of motion for the fourier-transformed
field, multiplied by $\widetilde{a}(\vec{q},t)$. The
equations of motion for the field, like
those for $\delta \rho$, are obtained from
eqn (\ref{newton}) and $T^{\mu \nu}_{~;\nu}=0$.} result:
\beq
i\frac{\partial}{\partial t} |\widetilde{a}(\vec{q},t)|^2
\simeq 
4 \pi m G_N  \sum_k \frac{R^2(t)}{|\vec{k}|^2}
\delta\rho(\vec{k},t)
\left\{\widetilde{a}^*(\vec{q}+\vec{k},t)\widetilde{a}(\vec{q},t)
-\widetilde{a}^*(\vec{q},t) \widetilde{a}(\vec{q}-\vec{k},t)
\right\} ~~~.
\label{finzzz}
\eeq
SY describe the axions as a coherent state,
so it is unsurprising that their calculation
gives the same result as  the classical field equations,
because coherent states are constructed for
that purpose. In the understanding of the
authors, the quantum aspect of the SY result
is to identify 
$ |\widetilde{a}(\vec{q},t)|^2/ \sum_p |\widetilde{a}(\vec{p},t)|^2$ 
as a number density of axion particles\footnote{
This is because the distribution of $\hbar$s
in the Lagrangian is different, depending on whether
the classical limit should be fields or particles \cite{hbar}.
So to define the particle number of a classical
field configuration requires $\hbar$.}.

An important difference  is that the axion number density
$n(\vec{q},t)$ studied by SY is labelled by the axion momentum,
whereas density fluctuations $\delta \rho(\vec{k},t)$ are labelled 
by the momentum of the graviton which they exchange. Notice
that the same  dynamics should be included in  
$\frac{\partial}{\partial t} n(\vec{q},t)$ of SY 
and  the equation of fluctuation growth (eqn \ref{ddd}), because
they are both the result of $T^{\mu \nu}_{~~;\nu} = 0$ and
the Poisson equation.  To
the understanding of the authors, the SY calculation shows
that the rate for an axion to emit a graviton
of any wavelength is large (compared to $H$). However,
what the gravitons then do is unknown. Whereas the
solution of the equations of motion for density fluctuations
say that the gravitons cause the density fluctuations to grow.

Finally,  SY show that axions
do not have coherent gravitational
interactions with other particles, such as photons,
in the early Universe plasma. This is
good, because  it means { that} axions are
not heated to the photon temperature. 
However, the classical Einsteins Equations
say that gravity is universal, so that
density fluctuations in the axions are
subject to the gravitational attraction
of other fluids (at linear order in $G_N$). 
Indeed, our estimate of the gravitational
damping scale involves the density
fluctuation, irrespective of whether
{ it is} made of axions or other particles.
How can these two perspectives be consistent?
At order $G_N$,   the axions should  have gravitational
interactions  with the fluctuations
in the density of other particles, rather than
with the individual particles. That is,
to find the universal gravitational
attraction between hot  other particles and
the cold axions, one should describe
the other particles with   an ``effective Lagrangian''
at the scale of the graviton momentum.
 For instance, in the Closed Time Path formalism
of SY,
 the radiation plasma in the early Universe
can be described (in 2PI  formalism)
by its two-point function. Averaged over
short distances  $\sim T^{-1}$, the two point
function becomes a Wigner function, which
can be approximated as a Boltzmann phase space
distribution on the  scale of axion momenta. The
density fluctuations encoded in the temperature
variations of this Boltzmann distribution
are the density fluctuations which 
 interact  gravitationally with the axions at order
$G_N$.  We interpret that the axions
do not interact with individual hot photons,
which could destroy the axion condensate,
but rather, that  the axion interactions
 with the long range density fluctuations
in other particles  will grow density perturbations,
and could contribute to the
 axion dissipation.


\section{ Summary}
\label{sec:sum}

{ The question of interest for this paper is  whether gravitational
interactions can ``thermalise''  the  axions produced
via the misalignment mechanism. We reproduce earlier estimates of
the gravitational interaction rate of these axions, 
but do not confirm that it is a thermalisation rate. 
}

We discuss this issue in cosmology, prior to  the epoch of
non-linear structure formation, because gravitational interactions
can be treated in the linear approximation. We suppose
that the Peccei-Quinn phase transition occurred
after inflation, so when the axion mass turns
on at the QCD phase transition and the axion  field
starts to oscillate, the coherence length of
the field is of order the horizon. Equivalently,
the comoving momentum of the field (or of the
axion particles in the coherent state
that makes it up) is of order the
expansion rate $H_{QCD}$. The axions  should
form CDM, therefore the gravitational interactions
of the homogeneous component must drive expansion, and
the gravitational interactions of density fluctuations
should cause them to grow. The question is whether,
in addition, gravity
can ``thermalise'' these axions, and cause them to
form a Bose Einstein condensate as anticipated by
Sikivie and collaborators \cite{SY1,ESTY}.
{ This question is only relevant, if
the misalignment axions are not already a Bose
Einstein condensate, as discussed in section
\ref{sec:BEC}.
}

The field theory literature indicates  that   Bose Einstein
condensation can arise in non-equilibrium situations,
as well as in thermal equilibrium -- but that
entropy is not generated in the leading order classical 
solution of time-reversal-invariant equations.
Instead,  some fluctuations must be resummed to 
obtain a Bose Einstein condensate in a calculation. 
So in this paper, we attempt to identify and
``resum''  some gravitational interactions which are not
those driving the expansion or growing density perturbations.
In section \ref{ssec:anis}, we estimate the contribution
of metric fluctuations to the off-diagonal elements of
the stress-energy tensor $T^i_j$. These elements are
commonly neglected in calculating the evolution
of axion density perturbations, so we imagine that
we can resum these fluctuations. We do this in section 
 \ref{ssec:TmnW}, by equating the  $T^i_j$ of section
 \ref{ssec:anis} to the  $T^i_j$ of an imperfect fluid
in a homogeneous and isotropic Universe. This gives
an estimate for the ``gravitational viscosity'' of
the axion fluid. We find that this viscosity damps fluctuations
on distances smaller than the axion Jeans length $\sqrt{1/m_aH}$.
The damping scale is given in eqn (\ref{dissipation}).
In particular, fluctuations on the comoving scale
$H_{QCD}^{-1}$ are not damped during the cosmological
periods we consider. So we do not confirm the
interpretation of \cite{ESTY,SY} that axions migrate
to the zero mode (form a Bose Einstein condensate)
at a photon temperature $T_\gamma \sim$ keV,
due to ``gravitational thermalisation''. 
We can reproduce the gravitational
interaction  rate obtained by  \cite{ESTY,SY}, but it
is unclear  to us that this is a thermalisation
rate: some of the gravitons should be contributing
to the growth of density fluctuations.
Section \ref{sec:disc} discusses our estimates
and compares to the calculation of \cite{SY}.

\section*{Acknowledgements}

We thank J. Gascon, A. Strumia, P Sikivie,
S.  Theissen and C. Wetterich  for useful conversations.
We are very grateful to
T. Noumi, K. Saikawa,  R.  Sato, and M. Yamaguchi,
for discussions and for allowing us to see their
work in progress \cite{SYetal2}.
And we especially  thank Georg Raffelt, for
uncountable  discussions, suggestions and  comments,
as well as for careful reading of the manuscript. 
The project was
performed in the context of the Lyon Institute of Origins, 
grant ANR-10-LABX-66,
and acknowledges  partial support from
the EU  FP7 ITN invisibles (MC Actions, PITN-GA-2011-289442).

\noindent
{\bf Note added}: When this paper was nearing completion, appeared 
an interesting discussion \cite{Alford:2013ota} of the 
connection between the field theory and  fluid descriptions
of a BE condensate.


\section{ Appendix}
\label{app}

In this Appendix, we study whether gravity can redistribute the
momenta of  CDM axions in a homogeneous
and isotropic Universe
described  by Einsteins Equations.
We  consider a classical free scalar field,
that is, a coherent state, 
evolving in a Friedman-Robertson-Walker universe.
We suppose this to be an adequate description of
dark matter axions after the  QCD phase
transition, during linear structure formation.
So  from an S-matrix perspective, we
take the ``in'' states to be axion modes shortly after
the QCD phase
transition, and the ``out'' states prior to $z\sim 10$. We 
describe  the  CDM axions 
 as a coherent state of ``in-particles''. 
We then evaluate, in that state,
 the expectation value of the number operator of
``out-particles'', thereby obtaining their momentum
distribution. As expected, the physical momentum of
the modes redshifts, and the comoving momentum distribution
does not change at leading order.


\subsection{ The Calculation }
\label{}

 We take the starting time $t = 0$
for our study to be  a few Hubble times after the QCD phase transition,
when the axion dark matter can be described as a real free 
scalar field in a Friedmann-Robertson-Walker background, with metric
\beq
ds^2 =  g_{\mu\nu} dx^\mu dx^\nu = dt^2 - R^2(t) [ dx^2 +  dy^2 +  dz^2  ] 
\label{metrique2}
\eeq
We follow the evolution of axion dark matter until this description breaks down,
when structure formation becomes non-linear. We  approximate this time as 
$t \to \infty$.  During this period, the axion field 
$a(x)$  satisfies the equations of motion:
\beq
\ddot{a}+ 3H\dot{a}
- \frac{1}{R^2(t)}\partial_i \partial_i a + m^2 a = 0
\label{EoM}
\eeq
where $\dot{a} = \frac{\partial}{\partial t} a$.
Following \cite{Parker,BD}, the field can be expanded on a complete
set of orthogonal eigenmodes 
 $\{u_{\vec{k}}(x)\}$, which
are solutions of eqn (\ref{EoM}), and which correspond to
axion particles at $t = 0$. We follow the conventions of
 \cite{BD}, but with the metric of eqn (\ref{metrique2}).
  It is convenient to normalise the
eigenmodes in a box of physical volume $R^3(t) L^3$:
\beq
u^{in}_{\vec{k}}(t, \vec{x}) = \frac{1}{[R(t) L]^{3/2}}\chi^{in}(t)e^{i \vec{k}\cdot \vec{x}}
\label{in}
\eeq
 Notice
that solutions of the Klein Gordon equation (\ref{EoM})  are separable,
due to the homogeneity of FRW spacetime. 

In a second-quantised formalism, the axion field 
 operator $\hat{a}(x)$ can be expanded
in the usual way on (time-dependant) annihilation and creation operators 
which satisfy $[\hat{b}_{\vec{k}}(t), \hat{b}^\dagger_{\vec{q}}(t)] = \delta_{\vec{k},\vec{q}}$ 
\cite{Parker},
and which multiply the modes  $\{u_{\vec{k}}\}$. These annihilation operators
define the ``in'' vacuum. 
The classical  axion  field   can therefore be written
as a coherent state \cite{I+Z} of (non-relativistic)
axion particles:
\bea
|a (\vec{x},t) \rangle &=& \frac{1}{N}
\exp\left\{ \sum_{\vec{p}} a(\vec{p},t) \hat{b}_{\vec{p}}^\dagger \right\}
|0_{in} \rangle~~~.
\label{coherentstate}
\eea
where $N$ is a normalisation factor to ensure  
$\langle a | a \rangle = 1$ .
This state  describes the classical axion field:
  $ \langle a |\hat{a}(x)| a \rangle  = a(x)$. 

We wish to know  the spectrum of axions that this state
describes at $t \to \infty$. It is well-known that gravity can
 change momenta and create particles. Canonical examples
are momentum red-shifting in FRW cosmologies, and  
black hole radiation: in  a  the curved space-time
outside a black hole, the state with no particles at $t \to -\infty$
will  contain particles at  $t \to +\infty$. This can be described \cite{BD} 
by writing the in- vacuum  creation operators (or equivalently,
eigenmodes)  in terms of the out-vacuum operators 
using Bogolibov coefficients:
\beq
u^{out}_{\vec{k}} = \sum_{\vec{q}}\alpha_{\vec{k}\vec{q}} u^{in}_{\vec{q}} +  
\beta_{\vec{k}\vec{q}} u^{in,*}_{\vec{q}} ~~~.
\label{Bogcoeff}
\eeq
Recall that the $\beta$ coefficients, which parametrise the
overlap between positive and negative frequency modes, 
 describe particle creation by gravity.
In the axion case, we wish to know the momentum distribution
of ``out-state'' axions --- that is, axion particles at the end of
linear structure formation ---in  the coherent
 state of eqn (\ref{coherentstate}). This can be evaluated if we
know the Bogoliubov transformation between the ``in'' and ``out''
creation and annihilation operators, or equivalently, if one can 
express the ``out'' eigenmodes in terms of the ``in'' eigenmodes,
as in eqn (\ref{Bogcoeff}). 

The out eigenmodes can be written
\beq
u^{out}_{\vec{k}}(t, \vec{x}) = \frac{1}{[R(t) L]^{3/2}}\chi^{out}(t)e^{i \vec{k}\cdot \vec{x}}
\label{out}
\eeq
where $\chi^{out}(t)$ is a solution of 
$$
\frac{\partial^2}{\partial t^2} {\chi}^{out}  
+ \frac{|\vec{k}|^2}{R^2(t)} {\chi}^{out} + m^2 {\chi}^{out} = 0
$$
and chosen to describe axion particles at $t \to \infty$
(${\chi}^{in}$ is a solution of the same equation).  
The homogeneity of FRW spacetimes  means that
co-moving momentum $\vec{k}$ is conserved, 
or equivalently, the Bogoliubov coefficients
are diagonal in momentum space \cite{BD}:
\beq
\alpha_{\vec{k},\vec{q}} = ( u^{out}_{\vec{k}},  u^{in}_{\vec{q}})  \propto 
 \delta_{\vec{k},\vec{q}}~~,~~~
  \beta_{\vec{k},\vec{q}}  = -( u^{out}_{\vec{k}},  u^{in,*}_{\vec{q}})  \propto 
 \delta_{\vec{k},-\vec{q}}
\eeq 
so  the number operator for  axion particles at  $t\to \infty$
is
\beq
\hat{b}^{out \dagger}_{\vec{k}} \hat{b}^{out}_{\vec{k}} =
( \alpha_{\vec{k},\vec{k}} \hat{b}^{in \dagger}_{\vec{k}}  -
 \beta_{\vec{k},-\vec{k}} \hat{b}^{in}_{-\vec{k}}) 
( \alpha^*_{\vec{k},\vec{k}}  \hat{b}^{in}_{\vec{k}} - \beta^*_{\vec{k},-\vec{k}}
 \hat{b}^{in \dagger}_{-\vec{k}} ) ~~~.
\label{reponse}
\eeq 
This shows that the effect of gravity on axions, in an expanding FRW
Universe, is to redshift their momenta (and possibly create particles).
There is no indication, from this calculation, that gravity
modifies the co-moving momentum distribution of the axions.

The axion creation by gravity, encoded in the coefficients $\beta$,
is expected to be negligible  because $H \ll m_a$. It can be
estimated, following \cite{Parker}, by taking the lowest order
adiabatic approximation  
\beq
\chi(t) = \frac{1}{ \sqrt{2 \omega}}e^{i \int^t \omega dt'}
\label{adapprox}
\eeq
with $\omega^2 = |\vec{k}|^2/R^2 + m^2$.  For
$|\vec{k}|^2 \ll m^2$, 
we obtain \footnote{Specifically,
by  substituting (\ref{adapprox}) into  eqn (22) of
\cite{Parker}, then integrating eqn (26) of
\cite{Parker} neglecting $\beta$ under the integral.} 
\beq
|\beta_{\vec{k},-\vec{k}}| \ll \frac{H (t = 0)}{m_a} ~~~~,~~~ 
\alpha_{\vec{k},\vec{k}} \simeq 1
\eeq 
so gravitational particle production
 can be  neglected, as expected,  because $H (t = 0) \simeq H_{QCD}
\ll m_a$.  Setting $ \beta_{\vec{k},-\vec{k}} \to 0$ in eqn (\ref{reponse})
implies that the number of axion particles making up the
classical field, and their co-moving momentum distribution, 
are unchanged  in the expanding FRW Universe.

\end{document}